\documentclass[pra,amsmath,superscriptaddress,twocolumn,eqsecnum,nofootinbib]{revtex4-2}
\pdfoutput=1
\usepackage[utf8]{inputenc}
\usepackage{times}
\usepackage{microtype}
\usepackage{graphicx}
\usepackage{xcolor}
\usepackage{adjustbox}
\usepackage{amssymb}
\usepackage{etoolbox}
\usepackage{titlesec}

\DeclareMathOperator{\Tr}{Tr}
\newcommand{\pvec}[1]{\vec{#1}\mkern2mu\vphantom{#1}}

\makeatletter
\newcommand*{\defeq}{\mathrel{\rlap{%
			\raisebox{0.3ex}{$\m@th\cdot$}}%
		\raisebox{-0.3ex}{$\m@th\cdot$}}%
	=}
\newcommand*{\eqdef}{=\mathrel{\rlap{%
			\raisebox{0.3ex}{$\m@th\cdot$}}%
		\raisebox{-0.3ex}{$\m@th\cdot$}}%
}
\makeatother

\makeatletter
\g@addto@macro\bfseries{\boldmath}
\makeatother

\makeatletter
\def\thesubsection{\thesection.\arabic{subsection}}
\def\p@subsection{}
\makeatother

\titleformat{\section}[hang]
{\normalfont\bfseries\MakeUppercase}
{\thesection.}
{1em}
{\raggedright}[]

\titleformat{\subsection}[hang]
{\normalfont\bfseries}
{\thesubsection.}
{1em}
{\raggedright}[]

\usepackage{scalerel}
\usepackage{tikz}
\usetikzlibrary{svg.path}
\definecolor{orcidlogocol}{HTML}{A6CE39}
\tikzset{
	orcidlogo/.pic={
		\fill[orcidlogocol] svg{M256,128c0,70.7-57.3,128-128,128C57.3,256,0,198.7,0,128C0,57.3,57.3,0,128,0C198.7,0,256,57.3,256,128z};
		\fill[white] svg{M86.3,186.2H70.9V79.1h15.4v48.4V186.2z}
		svg{M108.9,79.1h41.6c39.6,0,57,28.3,57,53.6c0,27.5-21.5,53.6-56.8,53.6h-41.8V79.1z M124.3,172.4h24.5c34.9,0,42.9-26.5,42.9-39.7c0-21.5-13.7-39.7-43.7-39.7h-23.7V172.4z}
		svg{M88.7,56.8c0,5.5-4.5,10.1-10.1,10.1c-5.6,0-10.1-4.6-10.1-10.1c0-5.6,4.5-10.1,10.1-10.1C84.2,46.7,88.7,51.3,88.7,56.8z};
	}
}
\newcommand\orcidlink[1]{\href{https://orcid.org/#1}{\mbox{\scalerel*{
				\begin{tikzpicture}[yscale=-1,transform shape]
					\pic{orcidlogo};
				\end{tikzpicture}
			}{X}}}}
		
\usepackage{url,hyperref}
\hypersetup{colorlinks,linkcolor={blue!55!black},citecolor={red!50!black},urlcolor={blue!45!black},breaklinks=true}

\begin{document}
	
\title{Probability vector representation of the Schrödinger equation and Leggett-Garg-type experiments}
	
\author{Masahiro Hotta}
\email{hotta@tuhep.phys.tohoku.ac.jp}
\affiliation{Department of Physics, Tohoku University, Sendai 980-8578, Japan}
\affiliation{Leung Center for Cosmology and Particle Astrophysics, National Taiwan University, Taipei 10617, Taiwan (R.O.C.)}
	
\author{Sebastian Murk\orcidlink{0000-0001-7296-0420}}
\email{sebastian.murk@oist.jp}
\affiliation{Quantum Gravity Unit, Okinawa Institute of Science and Technology, 1919-1 Tancha, Onna-son, Okinawa 904-0495, Japan}
	
\begin{abstract}
	Leggett-Garg inequalities place bounds on the temporal correlations of a system based on the principles of macroscopic realism \textit{per se} and noninvasive measurability. Their conventional formulation relies on the ensemble-averaged products of observables measured at different instants of time. However, a complete description that enables a precise understanding and captures all physically relevant features requires the study of probability distributions associated with noncommuting observables. In this article, we propose a scheme to describe the dynamics of generic $N$-level quantum systems (``qudits'') via a probability vector representation of the Schrödinger equation and define a precise notion of no-signaling in time (NSIT) for the probability distributions of noncommuting observables. This provides a systematic way of identifying the interferences responsible for nonclassical behavior. In addition, we introduce an interference witness measure to quantify violations of NSIT for arbitrary general probabilistic states. For single-qubit systems, we pinpoint the pivotal relation that establishes a connection between the disturbance of observables incurred during a measurement and the resulting NSIT violation. For large-$N$ systems where a manual determination is infeasible, the classification of states as either NSIT-conforming or NSIT-violating may be performed by a machine learning algorithm. We present a proof-of-principle implementation of such an algorithm in which the classifier function is prepared via supervised learning using pseudorandomly generated training data sets composed of states whose corresponding classifications are known \textit{a priori}.
\end{abstract}
	
\maketitle
	
\section{Introduction}
The distinction between classical and quantum phenomena has garnered considerable attention over the years. In contrast to the deterministic nature of classical mechanics, which accurately describes physical events on the macroscopic scales we experience in our day-to-day lives, quantum mechanics is fundamentally nondeterministic, and its precise role in the emergence of macroscopic phenomena is yet to be fully understood. To examine the breakdown of quantum coherence, Leggett and Garg devised an idealized experimental bound founded on the principles of macroscopic realism \textit{per se} (MRPS) and noninvasive measurability (NIM) \cite{LG:85}.\footnote{Our nomenclature in this article will refer explicitly to MRPS and NIM, and preclude the possibility of future measurements affecting measurement outcomes determined in the past, which is sometimes referred to as backwards causation (e.g.\ Ref.~\cite{K:16}) or induction (see, for instance, Refs.~\cite{KB:13,H:19} for possible definitions). In contrast to the original 1985 article by Leggett and Garg \cite{LG:85}, in the parlance of modern literature macroscopic realism (MR) is often defined as encompassing MRPS, NIM, and induction.}\ MRPS posits that physical properties of macroscopic systems exist independent of our observation, i.e., measurements on macroscopic systems merely reveal stable preexisting values. In other words, the moon is there even if nobody looks \cite{M:85}. In a trivial extension of quantum mechanics to large scales, macroscopic objects like Schrödinger's cat are described by a superposition of distinct states, and MRPS is broken. A more general concept of realism within hidden variable theories encompasses MRPS as a subset. NIM postulates that the measurement process has no bearing on the state of the system being measured, i.e., there is no backreaction of the measurement on the subsequent system dynamics.
	
Using the framework of general probabilistic theories \cite{P:23}, we consider a nonquantum system $S$ which has a corresponding $N$-dimensional quantum system $S_{\text{QM}}$ that shares the same observables. Then, there exists a complete set of $N^2 -1$ observables $\lbrace Q, \bar{Q}_1, \cdots, \bar{Q}_{N^2-2} \rbrace$ for $S_{\text{QM}}$ that uniquely determine the quantum density operator $\hat{\rho}$ of $S_{\text{QM}}$ via quantum tomography. However, in quantum mechanics the operators associated with these observables do in general not commute. The observables are assumed to be observed in $S$, at least when they are measured at distinct points in time. In noninvasive measurements of an observable $Q$ of the system $S$ (not $S_{\text{QM}}$), the probability distributions of all other observables $\bar{Q}_n$ remain unchanged while the initial probability distribution of $Q$ collapses into a more sharply peaked distribution [see Sec.~\ref{sec:PVR.SE}, Eqs.~\eqref{eq:SVcollapse.explicit} and \eqref{eq:SVcollapse}]. Such measurements do not exist in quantum mechanics, but may be allowed in more general theories like hidden variable theories. We regard the collapse of the initial probability distribution of $Q$ as a mere knowledge update about $Q$, not as a disturbance against the fundamental degrees of freedom of $S$, including $Q$. This interpretation of updating information through measurement without causing a disturbance aligns with the standard approach for macroscopic objects in classical statistical mechanics.\footnote{For an explicit example of noninvasive measurements in the classical theory, see Sec.~\ref{sec:SQ}, Eqs.~\eqref{eq:NIMC} and \eqref{eq:NIMC-2}.} 
	
Predicated on MRPS and NIM, experimentally testable inequalities of the form derived in Ref.~\cite{LG:85} [``Leggett-Garg inequalities'' (LGIs)]\footnote{See Ref.~\cite{ELN:rev:14} for a topical review.}\ bound the temporal correlations of a system in sequential measurements of observables. This is similar in spirit to the Bell \cite{B:64} and Clauser-Horne-Shimony-Holt (CHSH) inequalities \cite{CHSH:69}, which place bounds on the correlations in measurements of spatially separated systems based on the principles of realism and locality \cite{CK:16}.\footnote{Note that the experimental setting for testing LGIs is temporal: multiple measurements are performed within the same spacetime region enclosed by a single light cone. This is different from the spatially disconnected setting of Bell and CHSH inequalities, in which simultaneous measurements are performed in causally disconnected spacetime regions. Table~1 of Ref.~\cite{KB:13} compares the conceptual relationships in Bell- and Leggett-Garg-type inequalities.}\ A naive extrapolation of quantum mechanics to the macroscopic regime violates both types of inequalities. Reciprocally, the dynamics of a system that violates either LGIs or Bell/CHSH-type inequalities cannot be understood within the framework of traditional classical mechanics \cite{AGR:81,H:15}. 
	
Various proposals amenable to experimental verification of LGIs have been explored, including but not limited to quasiprobabilistic approaches \cite{H:16,HBLO:19,MNY:22}, continuous variable versions \cite{BHM:18}, and using expanded data sets obtained from finer-grained measurements \cite{MHL:21}. Violations of LGIs have been confirmed in numerous experiments involving different physical systems and using different types of measurements \cite{G:11,R:15,K:16,F:16,M:19} (see also Table 1 of Ref.~\cite{ELN:rev:14}).
	
Nevertheless, the precise scale up to which we can detect the quantumness of macroscopic objects in experiments remains elusive. LGIs are one of the principal tools to investigate how quantum coherence in the form of superpositions and/or entanglement is lost in the macroscopic realm \cite{WM:12,W:17}. In addition, the possibility of using LGIs to probe the quantumness of gravity through gravitationally induced violations has recently been put forward \cite{MNY:22}. 

In this article, we consider the no-signaling in time (NSIT) condition formulated in Refs.~\cite{KB:13,CK:15}, which posits that any measurements performed at $\tilde{t} < t$ do not change the outcome statistics of a later measurement at $t$, i.e., the probability $\mathcal{P}(Q=q,t)$ to observe the measurement outcome $q$ of an observable $Q$ at time $t$ is independent of whether or not measurements of arbitrary observables $\tilde{Q}$ have been performed at earlier times $\tilde{t}$, that is,  
\begin{align}
	\mathcal{P}(Q,t) - \mathcal{P}(Q,t)_{\vert (\tilde{Q}, \tilde{t})} = 0 , 
	\label{eq:def:NSIT}
\end{align}	
where the subscript $\vert (\tilde{Q}, \tilde{t})$ indicates that a measurement of the observable $\tilde{Q}$ has been performed at time $\tilde{t}<t$.
NSIT is a necessary mathematical condition for macroscopic realism (MR). While MR implies both LGIs and NSIT, the latter on its own is in general insufficient to derive LGIs. The NSIT condition is a strong coherence witness condition that is only satisfied if there is no interference. On the other hand, LGIs may still hold in the presence of moderate interference effects. In this sense, the NSIT condition is more stringent than the (non)classicality bounds imposed by LGIs. A comprehensive discussion of this aspect is provided in Ref.~\cite{H:17}. It is worth noting that --- while quantum mechanics generally violates NSIT due to interference terms --- it does obey the analogous no-signaling condition implied by local realism tested in Bell/CHSH inequalities [cf.\ Eq.~(4) and Tab.~1 of Ref.~\cite{KB:13}].

We also note that the hidden variable theory for a single qubit proposed by Bell \cite{B:64} precisely reproduces the results of quantum mechanics by violating the NSIT condition. In this theory, all expectation values $\langle \sigma_a \rangle_{\text{HV}}$ of the spin component $\sigma_a$ with $a \in \lbrace x,y,z \rbrace$ are allowed as long as $\langle \sigma_x \rangle^2_{\text{HV}} + \langle \sigma_y \rangle^2_{\text{HV}} + \langle \sigma_z \rangle^2_{\text{HV}} \leqslant 1$ holds. Then, it is possible to define density matrices via the following tomography relation:
\begin{align}
	\hat{\rho}_{\text{HV}} \defeq \frac{1}{2} \left( \hat{I}_2 + \langle \sigma_x \rangle_{\text{HV}} \hat{\sigma}_x + \langle \sigma_y \rangle_{\text{HV}} \hat{\sigma}_y + \langle \sigma_z \rangle_{\text{HV}} \hat{\sigma}_z \right) , 
	\label{eq:Bdm}
\end{align}
where $\hat{I}_2$ denotes the two-dimensional identity matrix and $\hat{\sigma}_a$ the Pauli matrices. Among the states described by $\hat{\rho}_{\text{HV}}$, there exist pure states denoted by $\vert \psi_{\text{HV}} \rangle \langle \psi_{\text{HV}} \vert$, which also appear in quantum mechanics. Similarly, all pure states in quantum mechanics are shared in Bell's theory. Therefore, the quantum coherence of $\vert \psi_{\text{HV}} \rangle$ is reproduced by a superposition of two distinct states $\vert \pm_{\text{HV}} \rangle$ via $\vert \psi_{\text{HV}} \rangle = c_+ \vert +_{\text{HV}} \rangle + c_- \vert -_{\text{HV}} \rangle$ with complex coefficients $c_{\pm}$, even though this is a hidden variable theory. In such nonlocal realism (i.e., neither local realism nor MR) theories, the NSIT condition is violated in the same manner as in quantum mechanics due to the unavoidable backreaction of the measurement.

NSIT tests are primarily tests for noninvasiveness \cite{H:17,H:19}. In previous studies of LGIs, arguments related to NIM have relied solely on the expectation values and ensemble averages of temporal correlations of observables. However, these quantities are secondary objects, derived from the probability distributions of observables in actual experiments. To provide a clearer and more fundamental description of quantum-mechanical interferences, we introduce a probability vector representation of the Schrödinger equation in this article. The notion of NSIT is unambiguously defined using the probability distributions in this representation. Based on this formalism, we precisely quantify the distinction between quantum mechanics and other theories in which NSIT holds. This ultimately allows us to better understand how fundamental differences between quantum mechanics and NSIT-compatible theories arise.
	
The remainder of this article is organized as follows: In Sec.~\ref{sec:PVR.SE}, we review mathematical preliminaries and derive the probability vector representation of the Schrödinger equation [Eq.~\eqref{eq:SchrEq.FPform}] which describes the evolution of a quantum system in terms of the probability distributions associated with its observables. In Sec.~\ref{sec:SQ}, we highlight important differences between classical [Subsec.~\ref{sec:SQ:subsec:CD}] and quantum [Subsec.~\ref{sec:SQ:subsec:QD}] dynamics based on the description of a single-qubit system and introduce an interference witness measure to quantify potential violations of NSIT [Eq.~\eqref{eq:SQ.gamma.a}] caused by the backreaction of a measurement [Eq.~\eqref{eq:SQ.Delta.def}] and their interrelation [Eq.~\eqref{eq:DeltaMeasure}]. The generalization to generic $N$-level quantum systems (``qudits'') is covered in Sec.~\ref{sec:NlvlSys}. In Sec.~\ref{sec:ML}, we outline how the classification of general probabilistic states as either NSIT-conforming or NSIT-violating could be performed by a machine learning algorithm in the case of very large $N$ (where a manual evaluation becomes infeasible in practice) and present a minimalistic proof-of-principle implementation. Lastly, we summarize our results and discuss their physical implications (Sec.~\ref{sec:C}).

\section{Probability vector Representation \newline of the Schrödinger Equation} \label{sec:PVR.SE}
\begin{figure*}[!htbp]
	\resizebox{0.925\linewidth}{!}{ 
		\includegraphics[scale=1.00]{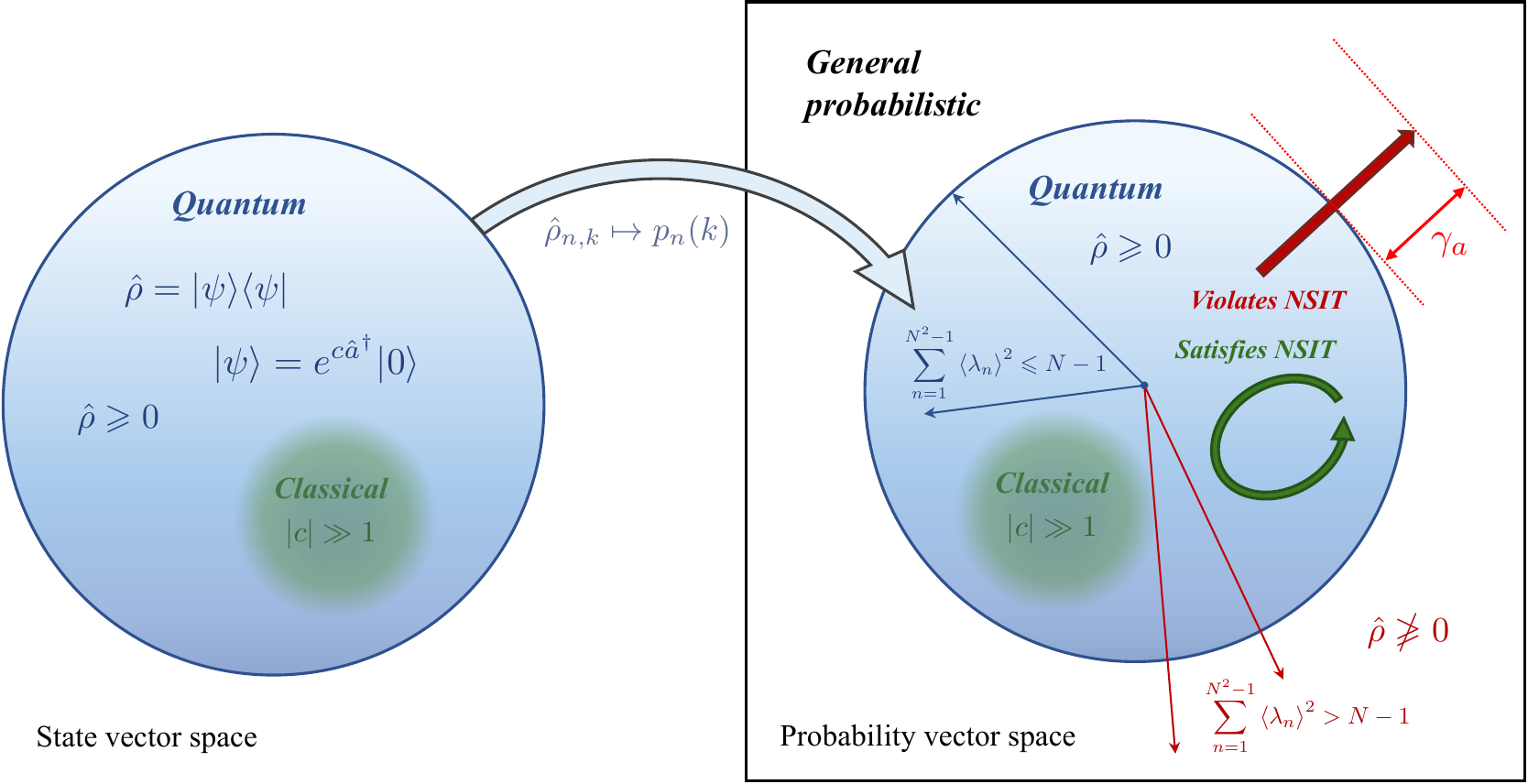}
	}
	\caption{Schematic illustration of the state vector space underlying the standard formalism of quantum mechanics (left) and the probability vector space employed in this article (right). There is a one-to-one correspondence between quantum states described by density operators (left) and their corresponding probability vector representation (right). A subset of the state vectors and their corresponding probability vectors (namely those with sufficiently large coherence amplitudes $\vert c \vert \gg 1$) display classical behavior (indicated by the green sphere nested within the blue quantum dynamics sphere). Here, $\hat{a}^\dagger$ denotes the raising/creation operator, and for $\vert c \vert \gg 1$ the expression $\vert \psi \rangle = e^{c \hat{a}^\dagger} \vert 0 \rangle$ describes a coherent state. A postmeasurement time evolution that does not (does) exhibit interference and therefore satisfies (violates) the NSIT condition [Eq.~\eqref{eq:def:NSIT}] is indicated by the thick green (red) arrow. The witness measure $\gamma_a$ [Eq.~\eqref{eq:SQ.gamma.a}] quantifies the severity of the NSIT violation, corresponding to the distance from the blue quantum dynamics sphere whose radius is determined by the Bloch sphere inequality [Eq.~\eqref{eq:Nsys.BlochSphere.Ineq}].} 
	\label{fig:StateSpace}
\end{figure*}

Our objective in this section is to derive a probability vector representation for quantum dynamics that is suitable for investigating violations of the NSIT condition [Eq.~\eqref{eq:def:NSIT}]. In the probability vector representation devised in this article, the quantum dynamics and quantum states are embedded into a more general probabilistic dynamics framework. A probability vector is a collection of probabilities associated with a finite number of physical observables. As illustrated in Fig.~\ref{fig:StateSpace}, the state vector space of quantum dynamics (left) corresponds to a subset of the general probabilistic probability vector state space (right). The quantum dynamics which is governed by a Hamiltonian corresponds to a stochastic evolution in the probability vector space. Measurements may result in unphysical postmeasurement states that do not satisfy the Bloch sphere inequality [cf.\ Eqs.~\eqref{eq:BlochSphere.Ineq}, \eqref{eq:BlochSphere.Ineq.EmProb}, and \eqref{eq:Nsys.BlochSphere.Ineq}] and thus have no quantum-mechanical equivalent/counterpart, i.e., there is no corresponding positive semidefinite density operator (which, in a slight abuse of notation, may be denoted as $\hat{\rho} \ngeqslant 0$) to describe the postmeasurement state. In this case, the stochastic evolution induced by the quantum Hamiltonian maps the probability vector into a vector that violates the NSIT condition (illustrated schematically by the thick red arrow exiting the Bloch sphere on the right-hand side of Fig.~\ref{fig:StateSpace}), which manifests itself in a time evolution in which some components of the probability vector take on negative values. In this section, we explicitly describe postmeasurement states that violate the NSIT condition and define an interference witness measure $\gamma_a$ [Eq.~\eqref{eq:SQ.gamma.a}] that quantifies how significant the deviation is from general probabilistic states whose probability vectors correspond to valid physical quantum states, i.e., those that satisfy the Bloch sphere inequality and have an associated positive semidefinite density operator $\hat{\rho} \geqslant 0$.

Let us first focus on the dynamics. The Schrödinger equation for an $N$-level system represented by the quantum state $\hat{\rho}(t)$ at time $t$ is given by
\begin{align}
	i \hbar \frac{d}{dt} \hat{\rho}(t) = \left[ \hat{H}, \hat{\rho}(t) \right] ,
		\label{eq:SchrEq}
\end{align}
where $\big[ \hat{A}, \hat{B} \big] \defeq \hat{A} \hat{B} - \hat{B} \hat{A}$ denotes the commutator of two operators $\hat{A}$ and $\hat{B}$. The generators $\hat{\lambda}_{n}$ of $\mathrm{SU(N)}$ satisfy
\begin{align}
	\hat{\lambda}_n^{\dag} = \hat{\lambda}_n , \quad \Tr \left[ \hat{\lambda}_n \right] = 0 , \quad \Tr \left[ \hat{\lambda}_n \hat{\lambda}_{n'} \right] = N \delta_{nn'} ,
	\label{eq:SU(N)gen.properties}
\end{align}
where $n,n' \in \lbrace 1, \dots, \mathrm{dim[SU(N)]} \rbrace$ with $\mathrm{dim[SU(N)]} = N^2-1$. The corresponding Lie algebra is given by
\begin{align}
	\left[ \hat{\lambda}_n, \hat{\lambda}_{n'} \right] = i \sum_{n''=1}^{N^2-1} \gamma_{nn'}^{n''} \hat{\lambda}_{n''} ,
\end{align}
where $\gamma_{nn'}^{n''}$ labels real-valued coefficients. Any $N$-level quantum state is decomposable in terms of the $\mathrm{SU(N)}$ generators $\hat{\lambda}_{n}$ via the so-called Bloch representation
\begin{align}
	\hat{\rho}(t) = \frac{1}{N} \left( \hat{I} + \sum_{n=1}^{N^2-1} \left \langle \lambda_n(t) \right \rangle \hat{\lambda}_n \right) , 
	\label{eq:BlochRep}
\end{align}
where $\hat{I}$ denotes the $N$-dimensional identity matrix, and the expectation values of the $\mathrm{SU(N)}$ generators are given by
\begin{align}
	\left \langle \lambda_n(t) \right \rangle = \Tr \left[ \hat{\lambda}_n \hat{\rho}(t) \right] .
	\label{eq:ExpVal}
\end{align}	
Their time derivatives are computed as
\begin{align}
	\frac{d}{dt} \left \langle \lambda_n(t) \right \rangle &= \Tr \left[ \hat{\lambda}_n \frac{d}{dt} \hat{\rho}(t) \right] = \frac{1}{i \hbar} \Tr \left[ \hat{\lambda}_n \left[ \hat{H}, \hat{\rho}(t) \right] \right] \nonumber \\
	&= \frac{1}{i \hbar} \Tr \left[ \hat{\rho}(t) \left[ \hat{\lambda}_n, \hat{H} \right] \right] .
	\label{eq:ExpVal.dt}
\end{align}
It is useful to introduce the real-valued coefficients
\begin{align}
	h_{nn'} &\defeq -\frac{i}{N} \Tr \left[ \hat{\lambda}_{n'} \left[ \hat{\lambda}_n, \hat{H} \right] \right] = \frac{i}{N} \Tr \left[ \hat{H} \left[ \hat{\lambda}_n, \hat{\lambda}_{n'} \right] \right] \nonumber \\
	&= -\frac{1}{N} \sum_{n''=1}^{N^2-1} \gamma_{nn'}^{n''} \Tr \left[ \hat{H} \hat{\lambda}_{n''} \right] .
\end{align}
One can then show that
\begin{align}
	\left[ \hat{\lambda}_n, \hat{H} \right] = i \sum_{n'=1}^{N^2-1} h_{nn'} \hat{\lambda}_{n'} .
\end{align}
The Schrödinger equation in the form of Eq.~\eqref{eq:SchrEq} can thus be recast in terms of the expectation values $\left \langle \lambda_n(t) \right \rangle$ as follows:
\begin{align}
	\frac{d}{dt} \left \langle \lambda_n(t) \right \rangle = \frac{1}{\hbar} \sum_{n'=1}^{N^2-1} h_{nn'} \left \langle \lambda_{n'}(t) \right \rangle .
	\label{eq:SchrEq.ExpValForm}
\end{align}
This is a generalization of the standard Bloch equation for a single qubit [$N=2$] to arbitrary $N$. The spectral decomposition of the $\mathrm{SU(N)}$ generators $\hat{\lambda}_n$ is given by
\begin{align}
	\hat{\lambda}_n = \sum_{k=1}^N \lambda_n(k) \hat{P}_n(k) ,
	\label{eq:SU(N)Gen.SpecDecomp}
\end{align}
where $\lambda_n(k)$ denote their eigenvalues and $\hat{P}_n(k)$ their projectors, respectively. The emergent probability of $\lambda_n(k)$ for the observable $\hat{\lambda}_n$ in the state $\hat{\rho}(t)$ is
\begin{align}
	p_n(k,t) = \Tr \left[ \hat{P}_n(k) \hat{\rho}(t) \right]
	\label{eq:EmProb}
\end{align}
and satisfies the normalization condition
\begin{align}
	\sum\limits_{k=1}^{N} p_n(k,t) = 1 .
	\label{eq:normalization}
\end{align}
The expectation values can be expressed in terms of their respective emergent probabilities, i.e.,
\begin{align}
	\left \langle \lambda_n(t) \right \rangle = \sum\limits_{k=1}^{N} \lambda_n(k) p_n(k,t) ,
	\label{eq:ExpVal.EmProb}
\end{align}
and their time derivatives are computed using
\begin{align}
	\frac{d}{dt} p_n(k,t) &= \Tr \left[ \hat{P}_n(k) \frac{d}{dt} \hat{\rho}(t) \right] = \frac{1}{i \hbar} \Tr \left[ \hat{P}_n(k) \left[ \hat{H}, \hat{\rho}(t) \right] \right] \nonumber \\
	&= \frac{1}{i \hbar} \Tr \left[ \hat{\rho}(t) \left[ \hat{P}_n(k), \hat{H} \right] \right] .
	\label{eq:dt.EmProb}
\end{align}
Expanding the right-hand side of this equation with respect to the generators $\hat{\lambda}_n$ yields
\begin{align}
	\frac{1}{i \hbar} \left[ \hat{P}_n(k), \hat{H} \right] = \sum\limits_{n'=1}^{N^2-1} K_{nn'}(k) \hat{\lambda}_{n'} , 
	\label{eq:dt.EmProb.RHSexpanded}
\end{align}
with coefficients $K_{nn'}(k)$ given by
\begin{align}
	K_{nn'}(k) = \frac{1}{N i \hbar} \Tr \left[ \hat{\lambda}_{n'} \left[ \hat{P}_n(k), \hat{H} \right] \right] .
\end{align}
Substituting the spectral decomposition of $\hat{\lambda}_{n'}$ [cf.\ Eq.~\eqref{eq:SU(N)Gen.SpecDecomp}] into Eq.~\eqref{eq:dt.EmProb.RHSexpanded}, we obtain
\begin{align}
	\frac{1}{i \hbar}\left[ \hat{P}_n(k), \hat{H} \right] = \sum_{n'=1}^{N^2-1} \sum_{k'=1}^{N} K_{nn'}(k) \lambda_{n'}(k') \hat{P}_{n'}(k') . 
\end{align}
Defining the coefficients
\begin{align}
	H_{nn'}(k,k') &\defeq \frac{\lambda_{n'}(k')}{N i \hbar} \Tr \left[  \hat{\lambda}_{n'} \left[ \hat{P}_n(k), \hat{H} \right] \right] , 
	\label{eq:Hnn'.coeff}
\end{align}
the following relation holds:
\begin{align}
	\frac{1}{i \hbar}\left[ \hat{P}_n(k), \hat{H} \right] = \sum_{n'=1}^{N^2-1} \sum_{k=1}^{N} H_{nn'}(k,k') \hat{P}_{n'}(k') . 
	\label{eq:Hnn'.rel}
\end{align}
Substitution of Eq.~\eqref{eq:Hnn'.rel} into Eq.~\eqref{eq:dt.EmProb} reveals that the Schrödinger equation [Eq.~\eqref{eq:SchrEq}] can be rewritten in the probability vector form [i.e., expressed in terms of the emergent probabilities $p_n(k,t)$] as follows:
\begin{align}
	\frac{d}{dt} p_n(k,t) = \sum_{n'=1}^{N^2-1} \sum_{k'=1}^{N} H_{nn'}(k,k') p_{n'}(k',t) . 
	\label{eq:SchrEq.FPform}
\end{align}
The initial condition of Eq.~\eqref{eq:SchrEq.FPform} is chosen such that the underlying probability distribution describes a valid quantum state. Therefore, $p_n(k,0) = \Tr \big[ \hat{P}_n(k) \hat{\rho}(0) \big]$ holds for an initial state described by the density matrix $\hat{\rho}(0)$. Using Eqs.~\eqref{eq:Hnn'.coeff}, \eqref{eq:SchrEq.FPform}, and the fact that $\sum_{k=1}^N \hat{P}_n(k) = \hat{I}$, it is straightforward to show that
\begin{align}
	\frac{d}{dt} \sum_{k=1}^N p_n(k,t) = 0 ,
\end{align}
and thus the normalization condition Eq.~\eqref{eq:normalization} holds at any time $t$:
\begin{align}
	\sum_{k=1}^N p_n(k,t) = \sum_{k=1}^N p_n(k) = 1 ,
\end{align}
where $p_n(k) \equiv p_n(k,0)$. Let $\vec{p}(t)$ represent the $N(N^2-1)$-dimensional probability vector
\begin{align}
	\vec{p}(t) = \left[
	\begin{array}[c]{c}
		p_{1}(1,t) \\
		\vdots \\
		p_{1}(N,t) \\
		\vdots \\
		p_{N^2-1}(1,t) \\
		\vdots \\
		p_{N^2-1}(N,t)
	\end{array}
	\right] .
	\label{eq:pVec}
\end{align}
Using the $N(N^2-1) \times N(N^2-1)$ matrix $H = \left[ H_{nn'} (k,k') \right]$ whose elements are prescribed by Eq.~\eqref{eq:Hnn'.coeff}, the solution of Eq.~\eqref{eq:SchrEq.FPform} is obtained as
\begin{align}
	\left[
	\begin{array}[c]{c}
		p_{1}(1,t) \\
		\vdots \\
		p_{1}(N,t) \\
		\vdots \\
		p_{N^2-1}(1,t) \\
		\vdots \\
		p_{N^2-1}(N,t)
	\end{array}
	\right] = \exp \left( t H \right) \left[
	\begin{array}[c]{c}
		p_{1}(1) \\
		\vdots \\
		p_{1}(N) \\
		\vdots \\
		p_{N^2-1}(1) \\
		\vdots \\
		p_{N^2-1}(N)
	\end{array}
	\right] ,
	\label{eq:SchrEq.FPform.Sol}
\end{align}
which can be rewritten in the series expansion form
\begin{align}
	\left[
	\begin{array}[c]{c}
		p_{1}(1,t) \\
		\vdots \\
		p_{1}(N,t) \\
		\vdots \\
		p_{N^2-1}(1,t) \\
		\vdots \\
		p_{N^2-1}(N,t)
	\end{array}
	\right] = \sum_{n=0}^{\infty} \frac{t^n}{n!} H^{n} \left[
	\begin{array}[c]{c}
		p_{1}(1) \\
		\vdots \\
		p_{1}(N) \\
		\vdots \\
		p_{N^2-1}(1) \\
		\vdots \\
		p_{N^2-1}(N)
	\end{array}
	\right] . 
	\label{eq:SchrEq.FPform.SolExpForm}
\end{align}
The numerical computation of $H^n$ does not require the exact diagonalization of $H$ and only takes a brief amount of time. Consequently, the right-hand side of Eq.~\eqref{eq:SchrEq.FPform.SolExpForm} can be calculated without facing significant obstacles even for sizable values of $N$.
	
For the measurement of an observable $\hat{\lambda}_n$ at time $t=0$, the expectation values $\left \langle \lambda_n \right \rangle$ of $\hat{\lambda}_n$ are determined by [cf.\ Eq.~\eqref{eq:ExpVal.EmProb}]
\begin{align}
	\left \langle \lambda_n \right \rangle = \sum_{k=1}^N \lambda_n(k) p_n(k) .
	\label{eq:ExpVal.idmeas}
\end{align}
If a noninvasive measurement of $\hat{\lambda}_1$ is performed and the result $\lambda_1(1)$ is obtained, the $\lambda_1$ sector of the probability vector $\vec{p}$ induces a collapse of the state vector such that $p'_1(k)_{1,1} = \delta_{k1}$, while the other sectors remain unaffected:
\begin{align}
	\resizebox{0.99\linewidth}{!}{$
		\vec{p} = \left[
		\begin{array}[c]{c}
			p_{1}(1) \\				
			p_{1}(2) \\
			\vdots \\
			p_{1}(N) \\
			p_{2}(1) \\
			\vdots \\
			p_{2}(N) \\
			\vdots \\
			p_{N^{2}-1}(1) \\
			\vdots \\
			p_{N^{2}-1}(N)
		\end{array}
		\right] 
		\Rightarrow 
		\pvec{p}'_{1,1} = \left[
		\begin{array}[c]{c}
			p_{1}^{\prime}(1)_{1,1} \\
			p_{1}^{\prime}(2)_{1,1} \\
			\vdots \\
			p_{1}^{\prime}(N)\,_{1,1} \\
			p_{2}^{\prime}(1)_{1,1} \\
			\vdots \\
			p_{2}^{\prime}(N)_{1,1} \\
			\vdots \\
			p_{N^{2}-1}^{\prime}(1)_{1,1} \\
			\vdots \\
			p_{N^{2}-1}^{\prime}(N)_{1,1}
		\end{array}
		\right] 
		= \left[
		\begin{array}[c]{c}
			1 \\
			0 \\
			\vdots \\
			0 \\
			p_{2}(1) \\
			\vdots \\
			p_{2}(N) \\
			\vdots \\
			p_{N^{2}-1}(1) \\
			\vdots \\
			p_{N^{2}-1}(N)
		\end{array}
		\right] .
		$}
	\label{eq:SVcollapse.explicit}
\end{align}	
This provides a precise definition of NIM for the underlying probability distribution. Tests of NSIT assess the effect of this state reduction in sequential measurements, i.e., measurements of the two-time (or three-time, etc.)\ probability. Analogously, one can define probability distributions $p'_{n'}(k')_{n,k}$ after a noninvasive measurement of $\hat{\lambda}_n$ at time $t=0$ results in the observation of the eigenvalue $\lambda_n(k)$. The $\lambda_{n}$ sector of $\vec{p}$ induces a collapse of the state vector, while the other sectors remain unaffected:
\begin{align}
	p'_{n'}(k')_{n,k} = \delta_{n' n} \delta_{k' k} + \left( 1 - \delta_{n'n} \right) p_{n'}(k') .
	\label{eq:SVcollapse}
\end{align}
However, it is not assured that the postmeasurement probability vector $\pvec{p}'_{n,k} = [p'_{n'}(k^\prime)_{n,k}]$ always represents a valid quantum state. The expectation values of $\hat{\lambda}_{n^\prime}$ are evaluated as
\begin{align}
	\left \langle \lambda_{n'} \right \rangle'_{n,k} = \sum_{k'=1}^N \lambda_{n'}(k') p'_{n'}(k')_{n,k} .
\end{align}
The postmeasurement density matrix is given by
\begin{align}
	\hat{\rho}'_{n,k} = \frac{1}{N} \left( \hat{I} + \sum_{n'=1}^{N^2-1} \left \langle \lambda_{n'} \right \rangle'_{n,k} \hat{\lambda}_{n'} \right) 
	\label{eq:FK.PMrho}
\end{align}
and could possess negative eigenvalues, which would imply that the corresponding operator is no longer positive semidefinite, $\hat{\rho}'_{n,k} \ngeqslant 0$. In this sense, the presence of negative eigenvalues signifies the violation of NSIT.
	
Next, for an arbitrary initial state $\hat{\rho}(0)$, let us introduce a witness measure $\gamma_{n,k}$ that quantifies the violation of the NSIT condition [Eq.~\eqref{eq:def:NSIT}] when a measurement of $\hat{\lambda}_{n}$ observes $\lambda_{n}(k)$. To this end, we first evaluate $p_{n'}(k')$ via $\Tr \big[ \hat{P}_{n'}(k') \hat{\rho}(0) \big]$ [cf.\ Eq.~\eqref{eq:EmProb}]. After the measurement, its $\lambda_n$ sector undergoes the following transition:
\begin{align}
	p_{n}(k') \; \Rightarrow \; p'_{n}(k') = \delta_{k'k} .
\end{align}
Then, the expectation value of $\hat{\lambda}_{n}$ is computed as
\begin{align}
	\left \langle \lambda_{n} \right \rangle'_{n,k} = \sum_{k'=1}^{N} \lambda_{n'}(k') p'_{n}(k') = \lambda_{n}(k) .
\end{align}
For sectors $n' \neq n$ on the other hand, the probabilities remain unchanged,
\begin{align}
	p_{n'}(k') \; \Rightarrow \; p'_{n'}(k') = p_{n'}(k') ,
\end{align}
and the expectation value of $\hat{\lambda}_{n' \neq n}$ is given by
\begin{align}
	\left \langle \lambda_{n'} \right \rangle'_{n,k} \equiv \left \langle \lambda_{n'} \right \rangle = \Tr \left[ \hat{\lambda}_{n'} \hat{\rho}(0) \right] .
\end{align}
For $\hat{\rho}(0)$ the pseudodensity matrix after the measurement can be written as\footnote{Note that dynamically evolving quantum systems that encode temporal correlations are necessarily represented by pseudodensity matrices \cite{FJV:16}.}
\begin{align}
	\hat{\rho}'_{n,k} = \frac{1}{N} \left( \hat{I} + \sum_{n'=1}^{N^2-1} \left \langle \lambda_{n'} \right \rangle'_{n,k} \hat{\lambda}_{n'} \right) .
\end{align}
More specifically, it can be expressed as
\begin{align}
	\hat{\rho}'_{n,k} = \frac{1}{N} \left( \hat{I} + \lambda_{n}(k) \hat{\lambda}_{n} + \sum_{n' \neq n} \Tr \left[ \hat{\lambda}_{n'} \hat{\rho}(0) \right] \hat{\lambda}_{n'} \right) . 
	\label{eq:pseudo.rho.afterMRM}
\end{align}
Let $p'_{m}(n,k)$ label the eigenvalues in the spectral decomposition of $\hat{\rho}'_{n,k}$, i.e.,
\begin{align}
	\hat{\rho}'_{n,k} = \sum_{m=1}^{N} p'_{m}(n,k) \hat{P}'_{m}(n,k) .
		\label{eq:FP.SpecDecomp}
\end{align}
The interference witness measure $\gamma(n,k)$ for $\hat{\rho}(0)$ is then defined as the sum of the absolute values of all negative eigenvalues:
\begin{align}
	\gamma_{n,k} \defeq \hspace*{-3.3mm} \sum_{p'_{m}(n,k) < 0} \hspace*{-3mm} \big\vert p'_{m}(n,k) \big\vert . 
	\label{eq:def.LGI.gamma}
\end{align}
As already alluded to in the caption of Fig.~\ref{fig:StateSpace}, the possible appearance of negative eigenvalues in the spectrum of the postmeasurement density operator is inherently connected to states described by probability vectors whose components are emergent probabilities that do not satisfy the corresponding Bloch sphere inequality [cf.\ Eq.~\eqref{eq:BlochSphere.Ineq.EmProb}], as will become evident in what follows, e.g., in the explicit examples considered in Eqs.~\eqref{eq:py.p1} and \eqref{eq:py.m1}.

The conclusion of this section warrants the following final remark: one can certainly verify the violation of NSIT by numerically diagonalizing $\hat{\rho}'_{n,k}$, identifying the negative eigenvalues in its spectrum, and then confirming that $\gamma_{n,k} > 0$. However, executing such a numerical diagonalization for large $N$ is a notably huge task that requires a significant amount of computational resources compared to the computation of $H^{n}$. Therefore, Eq.~\eqref{eq:SchrEq.FPform.SolExpForm} provides a much more efficient way of investigating violations of the NSIT condition in large-$N$ systems. Some of the $p_{n'}(k',t)$ values in Eq.~\eqref{eq:SchrEq.FPform.SolExpForm} can become negative at specific instances $t$ by taking  $p_{n'}(k') = \Tr \big[ \hat{P}_{n'}(k') \hat{\rho}'_{n,k} \big]$ and an adequate Hamiltonian $\hat{H}$. By solving this equation, one can identify the presence of negative components within $p_{n'}(k',t)$, which serves as an indicator of the NSIT violation for $\hat{\rho}(0)$ in the large-$N$ case.
	
\section{Single-Qubit Systems} \label{sec:SQ}
To account for the backreaction of quantum measurements on the probability distributions of observables for a single-qubit system, we first revisit the Bloch representation of quantum states. For a single qubit [$N=2$], the quantum state $\hat{\rho}$ is precisely specified by the expectation values $\langle \sigma_x \rangle$, $\langle \sigma_y \rangle$, $\langle \sigma_z \rangle$ of the three Pauli operators $\hat{\sigma}_x$, $\hat{\sigma}_y$, $\hat{\sigma}_z$ via
\begin{align}
	\hat{\rho} = \frac{1}{2} \left( \hat{I} + \langle \sigma_x \rangle \hat{\sigma}_x + \langle \sigma_y \rangle \hat{\sigma}_y + \langle \sigma_z \rangle \hat{\sigma}_z \right) . 
	\label{eq:SQ.BlochRep}
\end{align}
The state space is represented by a Bloch sphere [see Fig.~\ref{fig:SQSS}], which is defined by the inequality
\begin{align}
	\langle \sigma_x \rangle^2 + \langle \sigma_y \rangle^2 + \langle \sigma_z \rangle^2 \leqslant 1 .
	\label{eq:BlochSphere.Ineq}
\end{align}
This condition guarantees that all eigenvalues of $\hat{\rho}$ remain non-negative, which is commonly (again, in a slight abuse of notation) denoted as $\hat{\rho} \geqslant 0$. The emergent probabilities of the measurement outcomes $\pm 1$ for $\hat{\sigma}_a$ with $a \in \lbrace x,y,z \rbrace$ are computed as
\begin{align}
	p_a \left( \pm 1 \right) = \Tr \left[ \hat{P}_a(\pm 1) \hat{\rho} \right] ,
\end{align}
where the projection operators $\hat{P}_a(\pm 1)$ for $\hat{\sigma}_a$ are represented by $\hat{P}_a(\pm 1) \defeq \frac{1}{2} \big( \hat{I} \pm \hat{\sigma}_a \big)$. The Bloch sphere inequality of Eq.~\eqref{eq:BlochSphere.Ineq} can be rewritten in terms of the emergent probabilities as
\begin{align}
	\begin{aligned}
		& \big(p_{x}(+1)-p_{x}(-1)\big)^{2} + \big(p_{y}(+1)-p_{y}(-1)\big)^{2} \\
		& \hspace{14mm} + \big(p_{z}(+1)-p_{z}(-1)\big)^{2} \leqslant 1 .
	\end{aligned}
	\label{eq:BlochSphere.Ineq.EmProb}
\end{align}
To examine violations of NSIT for a single-qubit system described by $\hat{\rho}$, it is convenient to consider the six-dimensional [in general $N(N^2-1)$-dimensional] real vector of emergent probabilities given by
\begin{align}
	\vec{p} = \left[
	\begin{array}[c]{c}
		p_{x}(+1) \\
		p_{x}(-1) \\
		p_{y}(+1) \\
		p_{y}(-1) \\
		p_{z}(+1) \\
		p_{z}(-1)
	\end{array}
	\right] .
	\label{eq:SQ.EmProbVec}
\end{align}
The generic form of the time-independent Hamiltonian for a single qubit is given up to a constant by
\begin{align}
	\hat{H} = \frac{\hbar}{2} \left( B_x \hat{\sigma}_x + B_y \hat{\sigma}_y + B_z \hat{\sigma}_z \right) , 
	\label{eq:SQ.Hamiltonian}
\end{align}
where $B_a$ with $a \in \lbrace x,y,z \rbrace$ denotes real parameters. In this case, the Schrödinger equation [Eq.~\eqref{eq:SchrEq}] can be expressed through the ordinary Bloch equation [Eq.~\eqref{eq:SchrEq.ExpValForm}] as
\begin{align}
	\frac{d}{dt} \left[
	\begin{array}[c]{c}
		\langle \sigma_{x}(t) \rangle \\
		\langle \sigma_{y}(t) \rangle \\
		\langle \sigma_{z}(t) \rangle
	\end{array}
	\right] = \left[
	\begin{array}[c]{ccc}
		0 & B_{z} & -B_{y} \\
		-B_{z} & 0 & B_{x} \\
		B_{y} & -B_{x} & 0
	\end{array}
	\right] \left[
	\begin{array}[c]{c}
		\langle \sigma_{x}(t) \rangle \\
		\langle \sigma_{y}(t) \rangle \\
		\langle \sigma_{z}(t) \rangle
	\end{array}
	\right] .
\end{align}
Before considering the quantum probability vector representation, it is sensible to first revisit the analogous classical theory to ensure a comprehensive understanding of why the NSIT condition is always satisfied by classical dynamics.

\subsection{Classical Dynamics} \label{sec:SQ:subsec:CD}
The classical equation of motion of a spin vector $\vec{S} = \big[S_{x}(t), S_{y}(t), S_{z}(t)\big]$ is given by \cite{B:46}
\begin{align} 
	\frac{d}{dt} \left[
	\begin{array}[c]{c}
		S_{x}(t) \\
		S_{y}(t) \\
		S_{z}(t)
	\end{array}
	\right] = \left[
	\begin{array}[c]{ccc}
		0 & B_{z} & -B_{y} \\
		-B_{z} & 0 & B_{x} \\
		B_{y} & -B_{x} & 0
	\end{array}
	\right] \left[
	\begin{array}[c]{c}
		S_{x}(t) \\
		S_{y}(t) \\
		S_{z}(t)
	\end{array}
	\right] ,
\end{align}
where the initial condition is specified through the continuous real parameters $S_{0a}$ $\big[a \in \lbrace x,y,z \rbrace \big]$ as
\begin{align}
	\left[
	\begin{array}[c]{c}
		S_{x}(0) \\
		S_{y}(0) \\
		S_{z}(0)
	\end{array}
	\right] = \left[
	\begin{array}[c]{c}
		S_{0x} \\
		S_{0y} \\
		S_{0z}
	\end{array}
	\right] .
\end{align}
In the following discussion, let the spin vector with its initial conditions be denoted by
\begin{align}
	\left[
	\begin{array}[c]{c}
		S_{x}(t) \\
		S_{y}(t) \\
		S_{z}(t)
	\end{array}
	\right] = \left[
	\begin{array}[c]{c}
		S_{x}(S_{0x},S_{0y},S_{0z},t) \\
		S_{y}(S_{0x},S_{0y},S_{0z},t) \\
		S_{z}(S_{0x},S_{0y},S_{0z},t)
	\end{array}
	\right] .
\end{align}
Let $\rho_{0}\left(S_{0x}, S_{0y}, S_{0z}\right)$ be the classical probability distribution of the initial spin satisfying
\begin{align}
	\rho_{0} \left(S_{0x},S_{0y},S_{0z}\right) \geqslant 0 ,
\end{align}
and
\begin{align}
	\int \!\! \int \!\! \int \rho_{0}\left(S_{0x},S_{0y},S_{0z}\right) dS_{0x} dS_{0y} dS_{0z} = 1 .
\end{align}
The distribution $\rho \left(S_{x},S_{y},S_{z},t\right)$ at time $t$ is determined by
\begin{widetext}	
	\begin{align}
		\begin{aligned}
			\rho\left(S_{x},S_{y},S_{z},t\right) &= \int \!\! \int \!\! \int 
			\delta \big(S_{x}-S_{x}(S_{0x},S_{0y},S_{0z},t)\big) 
			\delta \big(S_{y}-S_{y}(S_{0x},S_{0y},S_{0z},t)\big) 
			\delta \big(S_{z}-S_{z}(S_{0x},S_{0y},S_{0z},t)\big) \\
			& \hspace{66.5mm} \times \rho_{0}\left(S_{0x},S_{0y},S_{0z}\right) dS_{0x} dS_{0y} dS_{0z} ,
		\end{aligned}
	\end{align}
	and satisfies the equation of motion
	\begin{align} 
		\frac{\partial}{\partial t} \rho\left( S_{x},S_{y},S_{z},t\right) = -\left(S_{x},S_{y},S_{z}\right) 
		\left[
		\begin{array}[c]{ccc}
			0 & B_{z} & -B_{y} \\
			-B_{z} & 0 & B_{x} \\
			B_{y} & -B_{x} & 0
		\end{array}
		\right] \left[
		\begin{array}[c]{c}
			\frac{\partial}{\partial S_{x}} \\
			\frac{\partial}{\partial S_{y}} \\
			\frac{\partial}{\partial S_{z}}
		\end{array}
		\right] 
		\rho\left( S_{x},S_{y},S_{z},t\right) .
		\label{eq:cl.EOM}
	\end{align}
\end{widetext}
Consider a measurement of $S_{a}$ $\big[a \in \lbrace x,y,z \rbrace \big]$ resulting in the observed spin $s=\pm1$ of $S_{a}$ at $t=0$. The probability of $s$ for $S_{a}$ is computed via
\begin{align}
	\bar{p}_{a,s} = \int \!\! \int \!\! \int \Theta \left(sS_{0a}\right) \rho_{0} \left( S_{0x},S_{0y},S_{0z}\right) dS_{0x} dS_{0y} dS_{0z} ,
\end{align}
where $\Theta(x)$ denotes the Heaviside step function 
\begin{align}
	\Theta(x) \defeq \begin{cases}
		\; 1 \quad \; \; x \geqslant 0 \\
		\; 0 \quad \; \; x < 0
	\end{cases} .
\end{align}
The probability distribution subsequent to the measurement at $t=0$ is described by
\begin{align}
	\rho_{a,s} \left(S_{0x},S_{0y},S_{0z}\right) = \frac{\Theta\left(sS_{0a}\right)}{\bar{p}_{a,s}} \rho_{0}\left(S_{0x},S_{0y},S_{0z}\right). 
	\label{eq:PD.afterMRM}
\end{align}
From Eqs.~\eqref{eq:cl.EOM} and \eqref{eq:PD.afterMRM}, one can immediately ascertain that no backreaction from the measurement influences the expectation value of any observable at a future time $t$. Since the relation
\begin{align}
	\begin{aligned}
		\rho\left(S_{x},S_{y},S_{z},t\right) &= \bar{p}_{a,+1} \rho_{a,+1}\left(S_{x},S_{y},S_{z},t\right) \\
		& \hspace{7.8mm} + \bar{p}_{a,-1} \rho_{a,-1}\left(S_{x},S_{y},S_{z},t\right) 
	\end{aligned}
	\label{eq:NIMC}
\end{align}
is satisfied at time $t$, the expectation value
\begin{align}
	& \bar{p}_{a,+1} \!\! \int \!\!\! \int \!\!\! \int \!\! O(S_{x},S_{y},S_{z}) \rho_{a,+1}\left(S_{x},S_{y},S_{z},t\right) dS_{x} dS_{y} dS_{z} \nonumber \\
	& \; + \bar{p}_{a,-1} \!\! \int \!\!\! \int \!\!\! \int \!\! O(S_{x},S_{y},S_{z}) \rho_{a,-1}\left(S_{x},S_{y},S_{z},t\right) dS_{x} dS_{y} dS_{z}
	\label{eq:NIMC-2}
\end{align}
of a physical observable $O(S_{x},S_{y},S_{z})$ at time $t$ with measurement matches the expectation value $\int \!\! \int \!\! \int \! O(S_{x},S_{y},S_{z}) \rho(S_{x},S_{y},S_{z},t) dS_{x} dS_{y} dS_{z}$ of that same observable without the measurement. This ensures the stability and predictability of the system even after measurements have been performed, thereby underscoring the classical nature of the described dynamics. Therefore, classical statistical mechanics is an example of an NSIT-compatible theory.
	
\subsection{Quantum Dynamics} \label{sec:SQ:subsec:QD}
\begin{figure}[!htbp]
	\resizebox{0.945\linewidth}{!}{ 
		\includegraphics[scale=1.00]{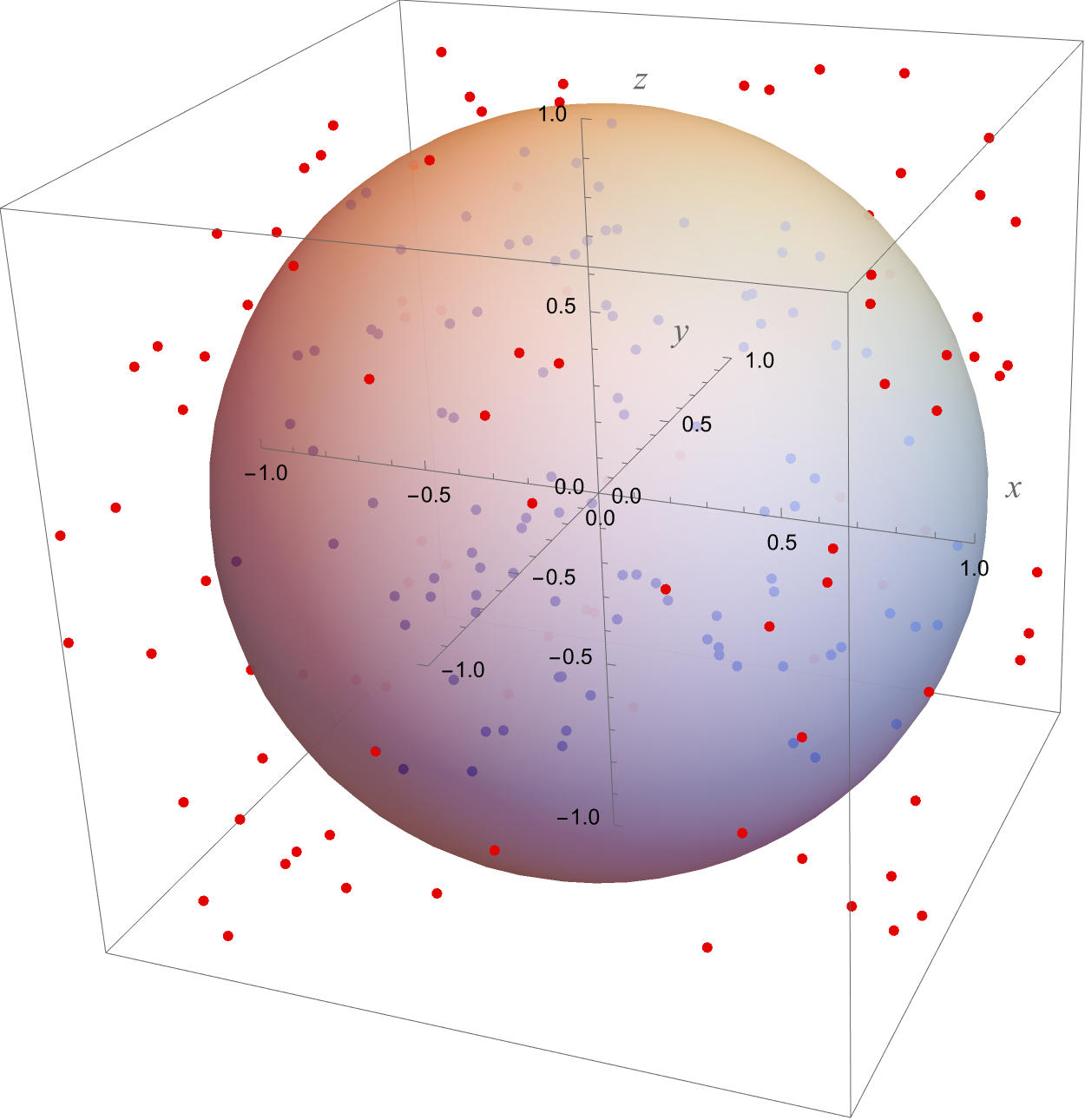}
	}
	\caption{Illustration of the state space of a single qubit [$N=2$]. A random sample of 100 states that satisfy the Bloch sphere inequality Eq.~\eqref{eq:BlochSphere.Ineq} is represented by the points in blue. Such points are necessarily located within the Bloch sphere. The red points correspond to a random sample of 100 states that satisfy Eq.~\eqref{eq:CSS}, but do not satisfy the Bloch sphere inequality Eq.~\eqref{eq:BlochSphere.Ineq}. Such points lie outside of the Bloch sphere, but within the general probabilistic state space defined by Eq.~\eqref{eq:CSS}, which is bounded by the cube and includes the Bloch sphere as a subset.} 
	\label{fig:SQSS}
\end{figure}

To consider measurements and violations of NSIT in quantum dynamical systems, we introduce the six-dimensional probability vector $\vec{p}_{\text{cl}}$ as a comparison measure for the above-described classical theory by defining the discrete spin variables $\sigma_{a}=\pm1$ as
\begin{align}
	\sigma_{x} &= \epsilon\left(S_{x}\right) = \Theta\left(S_{x}\right) - \Theta\left(-S_{x}\right) , \\
	\sigma_{y} &= \epsilon\left(S_{y}\right) = \Theta\left(S_{y}\right) - \Theta\left(-S_{y}\right) , \\
	\sigma_{z} &= \epsilon\left(S_{z}\right) = \Theta\left(S_{z}\right) - \Theta\left(-S_{z}\right) .
\end{align}
The classical probability vector $\vec{p}_{\text{cl}}(t)$ at time $t$ is then given by [cf.\ Eq.~\eqref{eq:SQ.EmProbVec}]
\begin{equation}
	\vec{p}_{\text{cl}}(t) = \left[
	\begin{array}[c]{c}
		\bar{p}_{x}(+1,t) \\
		\bar{p}_{x}(-1,t) \\
		\bar{p}_{y}(+1,t) \\
		\bar{p}_{y}(-1,t) \\
		\bar{p}_{z}(+1,t) \\
		\bar{p}_{z}(-1,t)
	\end{array}
	\right] = \left[
	\begin{array}[c]{c}
		\sum_{\sigma_{y},\sigma_{z}}p_{\text{cl}} (+1,\sigma_{y},\sigma_{z},t) \\
		\sum_{\sigma_{y},\sigma_{z}}p_{\text{cl}} (-1,\sigma_{y},\sigma_{z},t) \\
		\sum_{\sigma_{x},\sigma_{z}}p_{\text{cl}} (\sigma_{x},+1,\sigma_{z},t) \\
		\sum_{\sigma_{x},\sigma_{z}}p_{\text{cl}} (\sigma_{x},-1,\sigma_{z},t) \\
		\sum_{\sigma_{x},\sigma_{y}}p_{\text{cl}} (\sigma_{x},\sigma_{y},+1,t) \\
		\sum_{\sigma_{x},\sigma_{y}}p_{\text{cl}} (\sigma_{x},\sigma_{y},-1,t)
	\end{array}
	\right] ,
	\label{eq:clProbVec}
\end{equation}
where we use the bar to distinguish the classical probabilities $\bar{p}_a(s,t)$ from their quantum counterparts $p_a(s,t)$, and 
\begin{widetext}
	\begin{align}
		p_{\text{cl}} (\sigma_{x},\sigma_{y},\sigma_{z},t) = \int_{-\infty}^{\infty} \int_{-\infty}^{\infty} \int_{-\infty}^{\infty} \Theta(\sigma_{x}S_{x}) \Theta(\sigma_{y}S_{y}) \Theta(\sigma_{z}S_{z}) \rho(S_{x},S_{y},S_{z},t) dS_{x} dS_{y} dS_{z} .
	\end{align}
	By definition, each component $\bar{p}_{a}(s,t)$ remains non-negative at any arbitrary time $t$. Note that the general probabilistic state space defined by the set of points $\lbrace \langle \sigma_{x} \rangle_{\text{cl}}, \langle \sigma_{y} \rangle_{\text{cl}}, \langle \sigma_{z} \rangle_{\text{cl}} \rbrace$, where 
	\begin{align}
		\langle \sigma_{a} \rangle_{\text{cl}} \defeq (+1) \bar{p}_{a}(+1,t) + (-1) \bar{p}_{a}(-1,t) ,
		\label{eq:CSS}
	\end{align}
	is a cube with a side length of $2$, centered at the origin, with each side parallel to the $x$, $y$, and $z$ axis. Embedded within this cube is the Bloch sphere with a radius of $1$, making contact with the cube at its extremities, as illustrated in Fig.~\ref{fig:SQSS}. Any point that is located inside of the cube, yet not within the Bloch sphere (such as those indicated in red in Fig.~\ref{fig:SQSS}), corresponds to a probability distribution that does not align with our traditional understanding of quantum mechanics.
		
	Returning to quantum dynamics as delineated by Eq.~\eqref{eq:SchrEq.FPform}, the evolution of a single-qubit system is described by
	\begin{align}
		\frac{d}{dt} \left[
		\begin{array}[c]{c}
			p_{x}(+1,t) \\
			p_{x}(-1,t) \\
			p_{y}(+1,t) \\
			p_{y}(-1,t) \\
			p_{z}(+1,t) \\
			p_{z}(-1,t)
		\end{array}
		\right] =\left[
		\begin{array}[c]{cccccc}
			0 & 0 & -B_{z}/2 & B_{z}/2 & B_{y}/2 & -B_{y}/2 \\
			0 & 0 & B_{z}/2 & -B_{z}/2 & -B_{y}/2 & B_{y}/2 \\
			B_{z}/2 & -B_{z}/2 & 0 & 0 & -B_{x}/2 & B_{x}/2 \\
			-B_{z}/2 & B_{z}/2 & 0 & 0 & B_{x}/2 & -B_{x}/2 \\
			-B_{y}/2 & B_{y}/2 & B_{x}/2 & -B_{x}/2 & 0 & 0 \\
			B_{y}/2 & -B_{y}/2 & -B_{x}/2 & B_{x}/2 & 0 & 0
		\end{array}
		\right] \left[
		\begin{array}[c]{c}
			p_{x}(+1,t) \\
			p_{x}(-1,t) \\
			p_{y}(+1,t) \\
			p_{y}(-1,t) \\
			p_{z}(+1,t) \\
			p_{z}(-1,t)
		\end{array}
		\right] .
		\label{eq:SQ.QD.dt.ProbVec}
	\end{align}
\end{widetext}
In contrast to the probabilities $\bar{p}_{a}(s,t)$ in the case of classical dynamics described by Eq.~\eqref{eq:cl.EOM}, which are always non-negative, $p_{a}(s,t)$ can take on negative values in the quantum dynamics described by Eq.~\eqref{eq:SQ.QD.dt.ProbVec}, even when the same initial conditions are chosen for both equations. The presence of negative probability components $p_{a}(s,t)$ constitutes a direct indication for the violation of NSIT in quantum dynamics for the initial state with $\vec{p}_{\text{cl}}(0)$ in Eq.~\eqref{eq:clProbVec}. 

Next, let us reconsider the interference witness measure $\gamma_{n,k}$ introduced in Eq.~\eqref{eq:def.LGI.gamma} to quantify violations of the NSIT condition for a single qubit. Unlike the case of large $N$, we can easily determine its value for $N=2$. Let us assume that the initial state of the qubit is described by the probability vector $\vec{p}$ of Eq.~\eqref{eq:SQ.EmProbVec}. After performing a measurement of $\hat{\sigma}_{a}$ at $t=0$ and obtaining the result $s=\pm1$, we define a pseudo-density matrix denoted by $\hat{\rho}_{a,s}^{\prime}$. As per Eq.~\eqref{eq:pseudo.rho.afterMRM}, this matrix is described by the expression
\begin{align}
	\hat{\rho}'_{a,s} = \frac{1}{2} \left( \hat{I} + s \hat{\sigma}_{a} + \sum_{a' \neq a} \Tr \left[ \hat{\sigma}_{a'} \hat{\rho}(0) \right] \hat{\sigma}_{a'} \right) . 
	\label{eq:SQ.pseudo-rho}
\end{align}
Expanding Eq.~\eqref{eq:SQ.pseudo-rho} with respect to the identity matrix and the Pauli matrices [cf.\ Eq.~\eqref{eq:SQ.BlochRep}] as
\begin{equation}
	\hat{\rho}'_{a,s} = \frac{1}{2} \left( \hat{I} + \left\langle \sigma_{x} \right\rangle'_{a,s} \hat{\sigma}_{x} + \left\langle \sigma_{y} \right\rangle'_{a,s} \hat{\sigma}_{y} + \left\langle \sigma_{z} \right\rangle'_{a,s} \hat{\sigma}_{z} \right) 
\end{equation}
yields the following relations:
\begin{align}
	\left\langle \sigma_{a} \right\rangle'_{a,s} &= s, 
	\label{eq:SQ.BlochDecomp.Rel1} \\
	\left\langle \sigma_{b \neq a} \right\rangle'_{a,s} &= \Tr \left[ \hat{\sigma}_{b} \hat{\rho}(0) \right] ,
	\label{eq:SQ.BlochDecomp.Rel2}
\end{align}
where $a,b \in \lbrace x,y,z \rbrace$ here and in what follows. Note that $\hat{\rho}_{a,s}^{\prime}$ can possess a negative eigenvalue. Indeed, the two eigenvalues of the $2\times2$ matrix $\hat{\rho}'_{a,s}(0)$ are given explicitly by
\begin{align}
	\resizebox{0.99\linewidth}{!}{$ 
		p_{a,s_\pm} = \frac{1}{2} \left( 1 \pm \sqrt{ \left( \left\langle \sigma_{x} \right\rangle_{a,s}^{\prime} \right)^2 + \left( \left\langle \sigma_{y} \right\rangle_{a,s}^{\prime} \right)^2 + \left( \left\langle \sigma_{z} \right\rangle_{a,s}^{\prime} \right)^2} \right) .
		$}
\end{align}
Consequently, the witness measure $\gamma_{a,s}$ of Eq.~\eqref{eq:def.LGI.gamma} with $s=\pm1$ is evaluated as
\begin{equation}
	\resizebox{0.99\linewidth}{!}{$
		\gamma_{a,s} = \max \left\{ 0, \frac{1}{2} \left( \sqrt{\left( \left\langle \sigma_{x} \right\rangle'_{a,s} \right)^2 + \left( \left\langle \sigma_{y} \right\rangle'_{a,s} \right)^2 + \left( \left\langle \sigma_{z} \right\rangle'_{a,s} \right)^2} - 1 \right) \right\} . 
		$}
	\label{eq:SQ.MRBM}
\end{equation}
The value of $\gamma_{a,s}$ quantifies the NSIT violation and thus the extent to which the state under consideration differs from a physical postmeasurement state that satisfies the Bloch sphere inequality [see Fig.~\ref{fig:StateSpace}]. Since postmeasurement states that violate NSIT may not be realized in the experiment in general, $\gamma_{a,s}$ itself  does not qualify as a physical quantity. However, it is closely related to a physical quantity $\Delta$ that can be straightforwardly determined in experiments, as elucidated below. 
	
It follows from Eqs.~\eqref{eq:SQ.BlochDecomp.Rel1} and \eqref{eq:SQ.BlochDecomp.Rel2} that the squared expectation values of the Pauli matrices fulfill the following relation:
\begin{align}
	\begin{aligned}
		& \left( \left\langle \sigma_{x} \right\rangle'_{a,s} \right)^2 + \left( \left\langle \sigma_{y} \right\rangle'_{a,s} \right)^2 + \left( \left\langle \sigma_{z} \right\rangle'_{a,s} \right)^2 \\ 
		& \hspace*{10mm} = 1 + \sum\limits_{b \neq a} \Big( \! \Tr \left[ \hat{\sigma}_{b} \hat{\rho}(0) \right] \! \Big)^2 .
	\end{aligned}
	\label{eq:SQ.SqExpValRel}
\end{align}
Note that, since $\big(\! \left\langle \sigma_{x} \right\rangle'_{a,s} \!\big)^2 + \big(\! \left\langle \sigma_{y} \right\rangle'_{a,s} \!\big)^2 + \big(\! \left\langle \sigma_{z} \right\rangle'_{a,s} \!\big)^2 > 1$, the vector $\big[\! \left\langle \sigma_{x} \right\rangle'_{a,s}, \left\langle \sigma_{y} \right\rangle'_{a,s}, \left\langle \sigma_{z} \right\rangle'_{a,s} \!\big]$ lies outside of the Bloch sphere. From Eqs.~\eqref{eq:SQ.MRBM} and \eqref{eq:SQ.SqExpValRel}, $\gamma_{a,-s}$ is found to be equivalent to $\gamma_{a,s}$. In what follows, we therefore let $\gamma_{a}$ represent $\gamma_{a,s}$ given by
\begin{align}
	\gamma_{a} \equiv \gamma_{a,s} \defeq \frac{1}{2} \left( \sqrt{1 + \sum_{b \neq a} \big(\! \Tr \left[  \hat{\sigma}_{b} \hat{\rho}(0) \right] \!\big)^2} - 1 \right) .
	\label{eq:SQ.gamma.a}
\end{align}
The expectation values prior to the measurement are denoted by $\left\langle \sigma_{b}(0) \right\rangle = \Tr \left[ \hat{\sigma}_{b} \hat{\rho}(0) \right]$. Upon solving Eq.~\eqref{eq:SQ.gamma.a}, the following set of three equations is obtained:
\begin{align}
	\left\langle \sigma_{x}(0) \right\rangle^2 + \left\langle \sigma_{y}(0) \right\rangle^2 &= 4 \gamma_{z} (1+\gamma_{z}) , 
	\label{eq:SQ.ExpValRel.xy} \\
	\left\langle \sigma_{y}(0) \right\rangle^2 + \left\langle \sigma_{z}(0) \right\rangle^2 &= 4 \gamma_{x} (1+\gamma_{x}) , 
	\label{eq:SQ.ExpValRel.yz} \\
	\left\langle \sigma_{z}(0) \right\rangle^2 + \left\langle \sigma_{x}(0) \right\rangle^2 &= 4\gamma_{y} (1+\gamma_{y}) . 
	\label{eq:SQ.ExpValRel.zx}
\end{align}
The summation of Eqs.~\eqref{eq:SQ.ExpValRel.xy}--\eqref{eq:SQ.ExpValRel.zx} yields
\begin{equation}
	\left\langle \sigma_{x}(0) \right\rangle^2 + \left\langle \sigma_{y}(0) \right\rangle^2 + \left\langle \sigma_{z}(0) \right\rangle^2 = 2 \sum_{a} \gamma_{a} (1+\gamma_{a}) . 
	\label{eq:SQ.ExpValSum.gamma.a}
\end{equation}
In conjunction with the Bloch sphere condition Eq.~\eqref{eq:BlochSphere.Ineq} for the initial state $\hat{\rho}(0)$ [i.e., the left-hand side of Eq.~\eqref{eq:SQ.ExpValSum.gamma.a}], this relation establishes the following upper bound for the violation of the NSIT condition:
\begin{align}
	\sum_{a} \gamma_{a}(1+\gamma_{a}) \leqslant \frac{1}{2} .
	\label{eq:gamma.ub.MRB}
\end{align}
Based on Eq.~\eqref{eq:SQ.ExpValSum.gamma.a}, it is possible to derive an analogous inequality for the measured observables. Upon observing $\hat{\sigma}_{a}$, the averaged postmeasurement state is given by
\begin{align}
	\hat{\rho}_{a} = \sum_{s=\pm1} \hat{P}_{a,s} \hat{\rho}(t=0) \hat{P}_{a,s} ,
	\label{eq:SQ.PMstate}
\end{align}
where $\hat{P}_{a,s}= \vert s \rangle_{a} \langle s \vert_{a}$ are the projection operators of $\hat{\sigma}_{a}$ associated with the eigenvalues $s=\pm1$. The deviation of the expectation value of $\hat{\sigma}_{b}$ quantifies the backreaction of the measurement and is given by 
\begin{equation}
	\delta \sigma_b(a) = \left\langle \sigma_{b} \right\rangle^\prime_a - \left\langle \sigma_{b}(0) \right\rangle = \Tr \left[ \hat{\sigma}_b \hat{\rho}_a \right] - \Tr \left[  \hat{\sigma}_b \hat{\rho}(0) \right] .
\end{equation}
Based on this expression, we can introduce the measure $\Delta$ as follows:
\begin{align}
	\begin{aligned}
		\Delta &\defeq \frac{1}{3} \sum_{a} \sum_{b} \big( \delta\sigma_{b}(a) \big)^2 \\
		&= \frac{1}{3} \sum_{a} \sum_{b} \big(\! \left\langle \sigma_{b} \right\rangle'_{a} - \left\langle \sigma_{b}(0) \right\rangle \!\big)^2 . 
	\end{aligned}
	\label{eq:SQ.Delta.def}
\end{align}
As mentioned previously, this quantity can be determined by experiments. Using Eq.~\eqref{eq:SQ.PMstate} and the cyclic property of the matrix trace, we obtain
\begin{align}
	\delta \sigma_{b}(a) = \Tr \left[ \left( \sum_{s=\pm1} \hat{P}_{a,s} \hat{\sigma}_{b} \hat{P}_{a,s} - \hat{\sigma}_{b} \right) \hat{\rho}(0) \right] .
	\label{eq:SQ.delta.sigma.b}
\end{align}
The Pauli matrices satisfy the equation
\begin{align}
	\sum_{s=\pm1} \hat{P}_{a,s} \hat{\sigma}_{b\neq a} \hat{P}_{a,s} = 0 .
	\label{eq:P.sigma.P}
\end{align}	
From Eqs.~\eqref{eq:SQ.delta.sigma.b}--\eqref{eq:P.sigma.P}, it follows that
\begin{align}
	\delta \sigma_{b}(a) = \left( \delta_{ab} - 1 \right) \left\langle \sigma_{b}(0) \right\rangle .
\end{align}
This yields
\begin{align}
	\sum_{a} \sum_{b} \big(\delta\sigma_{b}(a)\big)^2 = 2 \left( \left\langle \sigma_{x}(0) \right\rangle^2 + \left\langle \sigma_{y}(0) \right\rangle^2 + \left\langle \sigma_{z}(0) \right\rangle^2 \right) . 
	\label{eq:delta.sigma.b.Sq}
\end{align}
From Eqs.~\eqref{eq:SQ.ExpValSum.gamma.a}, \eqref{eq:SQ.Delta.def}, and \eqref{eq:delta.sigma.b.Sq}, we obtain a useful formula relating the abstract quantity $\gamma_a$ to the experimentally observable quantity $\Delta$ given by
\begin{align}
	\Delta = \frac{4}{3} \sum_{a} \gamma_{a} \left( 1 + \gamma_{a} \right) .
	\label{eq:DeltaMeasure}
\end{align}
If $\Delta>0$, the single qubit in the initial state $\hat{\rho}(0)$ violates NSIT. Put simply, the relation in Eq.~\eqref{eq:DeltaMeasure} quantifies the extent to which the quantum world differs from a world without interference terms where $\gamma_{a}=0$ and NSIT is satisfied.
	
Next, we present examples of qubit quantum states to check the violation of NSIT. First, consider the maximally mixed state described by $\hat{\rho}(0)=\hat{I}/2$. The corresponding probability vector is given by
\begin{align}
	\vec{p} = 
	\left[
	\begin{array}[c]{c}
		p_{x}(+1,0) \\
		p_{x}(-1,0) \\
		p_{y}(+1,0) \\
		p_{y}(-1,0) \\
		p_{z}(+1,0) \\
		p_{z}(-1,0)
	\end{array}
	\right] = \left[
	\begin{array}[c]{c}
		\text{\textonehalf} \\
		\text{\textonehalf} \\
		\text{\textonehalf} \\
		\text{\textonehalf} \\
		\text{\textonehalf} \\
		\text{\textonehalf} 
	\end{array}
	\right] .
\end{align}
Prior to the measurement of $\hat{\sigma}_{a}$, all expectation values are null, i.e.\ $\left\langle \sigma_{a}(0) \right\rangle = p_a(+1,0) - p_a(-1,0) = \text{\textonehalf} - \text{\textonehalf} = 0 \; \forall a$. If the result $\sigma_{x}=+1$ is observed after a measurement of $\hat{\sigma}_{x}$ has been performed, the probability vector becomes
\begin{align}
	\pvec{p}'_{x,+1} = \left[
	\begin{array}[c]{c}
		p_{x}^{\prime}(+1)_{x,+1} \\
		p_{x}^{\prime}(-1)_{x,+1} \\
		p_{y}^{\prime}(+1)_{x,+1} \\
		p_{y}^{\prime}(-1)_{x,+1} \\
		p_{z}^{\prime}(+1)_{x,+1} \\
		p_{z}^{\prime}(-1)_{x,+1}
	\end{array}
	\right] = \left[
	\begin{array}[c]{c}
		1 \\
		0 \\
		\text{\textonehalf} \\
		\text{\textonehalf} \\
		\text{\textonehalf} \\
		\text{\textonehalf} 
	\end{array}
	\right] . 
\end{align}
The associated state $\hat{\rho}_{x,+1}^{\prime}$ is an eigenstate given by $\vert +_{x} \rangle \langle +_{x} \vert$ of $\hat{\sigma}_{x}$, thus establishing it as a quantum state that is compatible with NSIT. Hence, the interference witness measure for this particular state vanishes, i.e., $\gamma_{x,+1}=0$, and thus also $\Delta_{x,+1} =0$ by virtue of Eq.~\eqref{eq:DeltaMeasure}. Similarly, measurements of other components $\hat{\sigma}_{a}$ yield analogous quantum states. A single-qubit system whose initial state is described by $\hat{\rho}(0)=\hat{I}/2$ therefore satisfies the NSIT condition.
	
On the other hand, if a measurement of $\hat{\sigma}_{y}$ is performed on the initial state $\hat{\rho}(0) = \vert +_{x} \rangle \langle +_{x} \vert$ and the result $\sigma_{y}=+1$ is observed, the probability vector becomes
\begin{align}
	\pvec{p}'_{y,+1} = \left[
	\begin{array}[c]{c}
		p_{x}^{\prime}(+1)_{y,+1} \\
		p_{x}^{\prime}(-1)_{y,+1} \\
		p_{y}^{\prime}(+1)_{y,+1} \\
		p_{y}^{\prime}(-1)_{y,+1} \\
		p_{z}^{\prime}(+1)_{y,+1} \\
		p_{z}^{\prime}(-1)_{y,+1}
	\end{array}
	\right] = \left[
	\begin{array}[c]{c}
		1 \\ 
		0 \\ 
		1 \\ 
		0 \\
		\text{\textonehalf} \\
		\text{\textonehalf} 
	\end{array}
	\right] .
	\label{eq:py.p1}
\end{align}
Similarly, if the result $\sigma_{y}=-1$ is observed instead of $\sigma_{y}=+1$, the probability vector becomes	
\begin{align}
	\pvec{p}'_{y,-1} = \left[
	\begin{array}[c]{c}
		p_{x}^{\prime}(+1)_{y,-1} \\
		p_{x}^{\prime}(-1)_{y,-1} \\
		p_{y}^{\prime}(+1)_{y,-1} \\
		p_{y}^{\prime}(-1)_{y,-1} \\
		p_{z}^{\prime}(+1)_{y,-1} \\
		p_{z}^{\prime}(-1)_{y,-1}
	\end{array}
	\right] = \left[
	\begin{array}[c]{c}
		1 \\
		0 \\
		0 \\
		1 \\
		\text{\textonehalf} \\
		\text{\textonehalf} 
	\end{array}
	\right] . 
	\label{eq:py.m1}
\end{align}
Since $\pvec{p}'_{y,\pm1}$ does not adhere to the Bloch sphere condition prescribed by Eq.~\eqref{eq:BlochSphere.Ineq.EmProb}, the associated states $\hat{\rho}'_{y,\pm1}$ inevitably violate the NSIT condition. Indeed, the corresponding interference witness measure $\gamma_{y,\pm1}$ takes on positive values, namely
\begin{align}
	\gamma_{y,\pm1} = \frac{\sqrt{2} - 1}{2} .
\end{align}
In this case, under appropriate selection of $B_{a}$ in Eq.~\eqref{eq:SQ.Hamiltonian}, solving Eq.~\eqref{eq:SQ.QD.dt.ProbVec} reveals that some negative probability components $p_{a}(s,t)$ appear at a future time $t$. The negativity of the probability components $p_{a}(s,t)$ therefore serves as evidence for the violation of NSIT.
	
\section{N-Level Systems} \label{sec:NlvlSys}	
In a single-qubit system, every point contained within the Bloch sphere corresponds to a quantum state that can be realized in experiments. Analogous to the Pauli operators $\hat{\sigma}_a$ for $N=2$, one can introduce $N^2-1$ observables $\hat{\lambda}_{n}$ to describe the dynamics of generic $N$-level quantum systems [cf.\ Sec.~\ref{sec:PVR.SE}, Eqs.~\eqref{eq:SU(N)gen.properties}--\eqref{eq:ExpVal.dt}]. The $N$-level generalization of the Bloch sphere defining inequality Eq.~\eqref{eq:BlochSphere.Ineq} is
\begin{align}
	\sum_{n=1}^{N^2-1} \left\langle \lambda_{n} \right\rangle^2 \leqslant N-1 .
	\label{eq:Nsys.BlochSphere.Ineq}
\end{align}
Analogous to the single-qubit case with $N=2$, a saturation of this inequality corresponds to a pure quantum state. However, in contrast to the single-qubit case, a subset of the set of points $\lbrace \left\langle \lambda_{1} \right\rangle, \ldots, \left\langle \lambda_{N^2-1} \right\rangle \rbrace$ that satisfy the relation prescribed by Eq.~\eqref{eq:Nsys.BlochSphere.Ineq} does not describe valid quantum states \cite{K:03}. Consider, for instance, the quantum state described by $\hat{\rho}_{1} = \sum_{k} p_{k} \vert k \rangle \langle k \vert \geqslant 0$. Then, the vector $\vec{\lambda}_{1} = \left[ \left\langle \lambda_{1} \right\rangle _{1}, \ldots, \left\langle \lambda_{N^2-1} \right\rangle_{1} \right]$ defined by $\left\langle \lambda_{n} \right\rangle_{1} = \Tr \big[ \hat{\lambda}_{n} \hat{\rho}_{1} \big]$ provides the Bloch representation of $\hat{\rho}_{1}$, i.e.,
\begin{align}
	\hat{\rho}_{1} = \frac{1}{N} \hspace{0.5mm} \Bigg( \hat{I} + \sum_{n=1}^{N^{2}-1} \left\langle \lambda_{n} \right\rangle_{1} \hat{\lambda}_{n} \Bigg) .
\end{align}
Another quantum state $\hat{\rho}_{2}$ given by
\begin{align}
	\hat{\rho}_{2} = \frac{1}{N} \hspace{0.5mm} \Bigg( \hat{I} + \sum_{n=1}^{N^2-1} \left\langle \lambda_{n} \right\rangle_{2} \hat{\lambda}_{n} \Bigg)
\end{align}
should satisfy
\begin{align}
	\begin{aligned}
		\Tr \left[ \hat{\rho}_1 \hat{\rho}_2 \right] &= \sum_k p_k \langle k \vert \hat{\rho}_2 \vert k \rangle \\
		&= \frac{1}{N} \hspace{0.5mm} \Bigg( 1 + \sum_{n=1}^{N^2-1} \left\langle \lambda_n \right\rangle_1 \left\langle \lambda_n \right\rangle_2 \Bigg) \geqslant 0 .
	\end{aligned}
\end{align}
Hence, $\vec{\lambda}_{2} = \left[ \left\langle \lambda_{1} \right\rangle_{2}, \ldots, \left\langle \lambda_{N^2-1} \right\rangle_2 \right]$ obeys the following necessary condition:
\begin{align}
	\vec{\lambda}_{1} \cdot \vec{\lambda}_{2} = \sum_{n=1}^{N^2-1} \left\langle \lambda_{n} \right\rangle_{1} \left\langle \lambda_{n} \right\rangle_{2} \geqslant -1 . 
	\label{eq:lam1lam2prod}
\end{align}
It follows from Eq.~\eqref{eq:lam1lam2prod} that when $\hat{\rho}_{1}$ is a pure state satisfying $\sum_{n=1}^{N^2-1} \left\langle \lambda_{n} \right\rangle^2_1 =N-1$, a pure quantum state $\hat{\rho}_{2}$ which meets the condition $\left\langle \lambda_{n} \right\rangle_{2} = - \left\langle \lambda_{n} \right\rangle_{1}$ does not exist for $N \geqslant 3$. Consequently, the higher-dimensional generalization of the Bloch sphere inequality given in Eq.~\eqref{eq:Nsys.BlochSphere.Ineq} does not suffice to guarantee physically viable quantum states described by a positive semidefinite operator $\hat{\rho} \geqslant 0$. The implication here is that the state space characterized by $\left[ \left\langle \lambda_{1} \right\rangle, \ldots, \left\langle \lambda_{N^2-1} \right\rangle \right]$ is a rather intricate manifold. Therefore, for large values of $N$ it is in general quite difficult to determine whether a given vector $\left[ \left\langle \lambda_{1} \right\rangle, \ldots, \left\langle \lambda_{N^2-1} \right\rangle \right]$ corresponds to a valid quantum state or not since this requires the numerical diagonalization of $\hat{\rho}'(0)$. Similarly, for large $N$ it is a numerically difficult task to check if a given probability vector of the form of Eq.~\eqref{eq:pVec} describes a valid quantum state or not [recall that such a vector comprises $N(N^2-1)$ components]. In order to alleviate this difficulty, we propose to adopt a machine learning method as described in the following section.
	
\section{State Classification \newline with Machine Learning} \label{sec:ML}
The aim of our proposed machine learning method is to train an algorithm in the classification of general probabilistic states as either satisfying or violating the NSIT condition based on their associated probability vectors of the form given by Eq.~\eqref{eq:pVec}. Such vectors consist of $(N^2-1)$ probability tuples $[P_1, \ldots, P_{N^2-1}]$, each containing $N$ entries $[p_n(1,t), \ldots, p_n(N,t)]$ satisfying [cf.\ Eqs.~\eqref{eq:EmProb} and \eqref{eq:normalization}]
\begin{align}
	\sum\limits_k^N p_n(k,t) = 1 , \quad p_n(k,t)>0 \qquad \forall \; n,k
	\label{eq:ML.norm.pos}
\end{align}
with $n \in \lbrace 1, \ldots, N^2-1 \rbrace$ and $k \in \lbrace 1, \ldots, N \rbrace$. The first step in our approach is the generation of training data that can be used for supervised learning. Since our goal is to distinguish states that satisfy NSIT from those that violate it, we generate two distinct probability vector training data sets: the first containing exclusively vectors associated with states that satisfy NSIT, and the second containing exclusively vectors associated with states that violate NSIT.
	
\subsection{Training Data Generation}
\begin{figure*}[!htbp]
	\resizebox{0.95\linewidth}{!}{ 
		\includegraphics[scale=1.00]{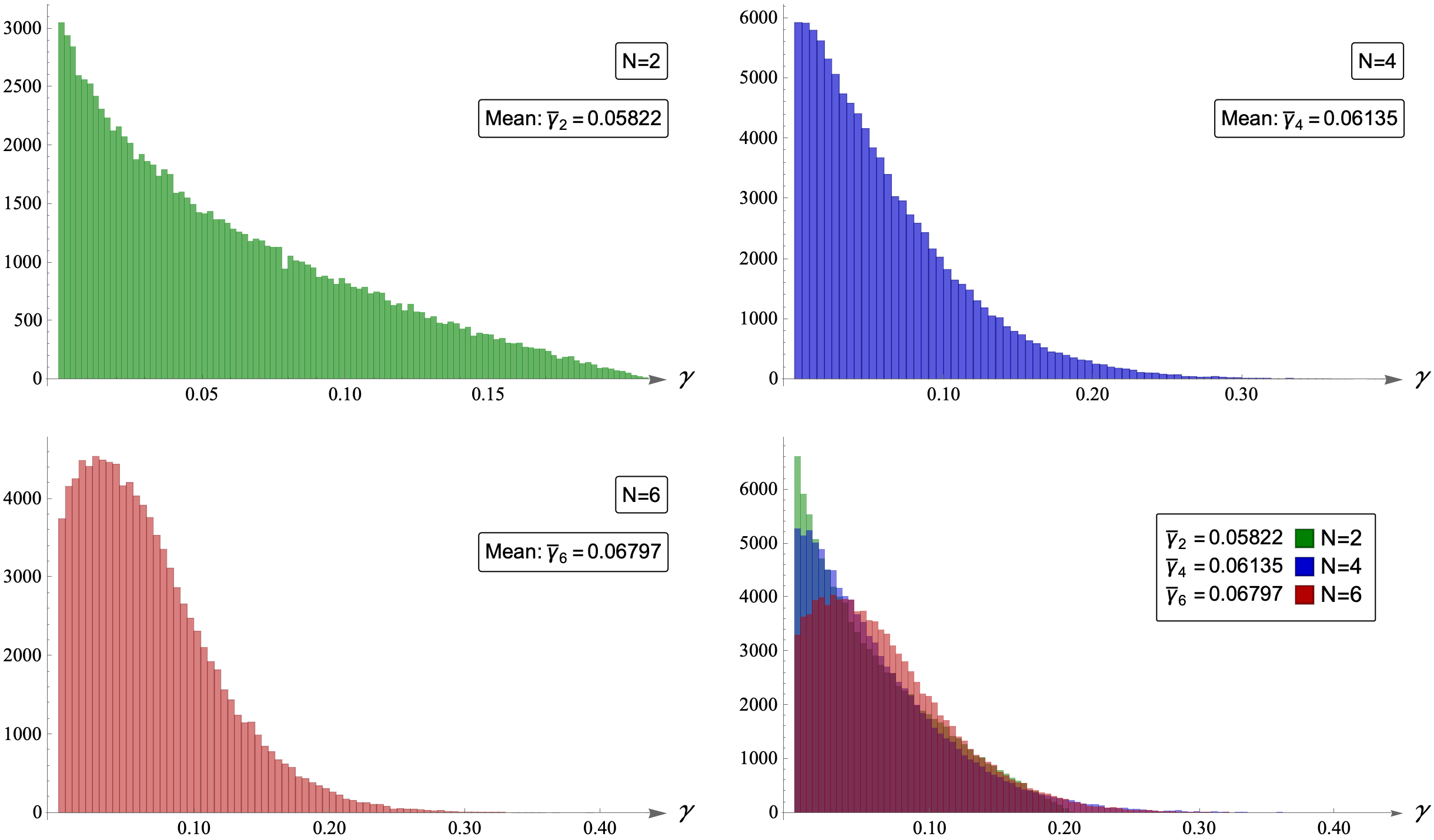}
	}
	\caption{Distribution of the interference witness measure $\gamma_{n,k}$ [Eq.~\eqref{eq:def.LGI.gamma}] for 100,000 pseudorandomly generated NSIT-violating states for $N \in \lbrace 2,4,6 \rbrace$ and their respective mean values $\bar{\gamma}_N = \frac{1}{100,000} \sum_i^{100,000} \gamma_{N_i}$. For the generated NSIT-violating training data sets, the observed mean value $\bar{\gamma}_N$ tends to increase with $N$ due to the fact that the pseudo-randomly generated $p'_m(n,k)$ that appear in the spectral decomposition of Eq.~\eqref{eq:FP.SpecDecomp} must also satisfy a normalization condition akin to the first relation in Eq.~\eqref{eq:ML.norm.pos}, i.e., they sum to 1 even in the presence of negative values. Therefore, in the current implementation underlying the pseudorandom generation of NSIT-violating states, there is more leeway for large $N$ to offset comparatively large negative values $p'_m(n,k)<0$ as the associated eigenvalue tuples have more entries available ($\sim \! N$, i.e., scaling linearly with $N$) to ensure a sum total of 1.} 
	\label{fig:gamma}
\end{figure*}

For any arbitrary $N \geqslant 2$, we work with the generalized Gell-Mann matrix basis (GGMMB) \cite{G:62}\footnote{Chapter 3 of Ref.~\cite{BK:08} provides an overview of the relevant properties.}\ and generate pseudodensity states according to the spectral decomposition of Eq.~\eqref{eq:FP.SpecDecomp}, where the projectors $\hat{P}'_m(n,k)$ are constructed from pseudorandomly generated $N$-dimensional vectors that are orthonormalized via the Gram-Schmidt process. The individual components of the probability vectors are then generated via Eq.~\eqref{eq:EmProb}, where this time the projectors $\hat{P}_n(k)$ are those associated with the elements of the GGMMB. The difference in the generation of probability vectors for the NSIT-conforming vs.\ the NSIT-violating data set lies in the pseudorandom generation of the coefficients $p'_m(n,k)$ of Eq.~\eqref{eq:FP.SpecDecomp}: while the normalization condition Eq.~\eqref{eq:normalization} is always satisfied for each of the $N$ probability tuples in both data sets, negative values $p'_m(n,k)<0$ are permitted in the generation of probability vectors associated with states that violate the NSIT condition to reflect the fact that the spectrum of the postmeasurement density matrix [cf.\ Eq.~\eqref{eq:FK.PMrho}] may contain negative eigenvalues and thus cannot describe a physically valid state. This may ultimately result in probability vectors with negative components $p_n(k,t)<0$. However, since this contradicts the second requirement stipulated by Eq.~\eqref{eq:ML.norm.pos}, such vectors are then discarded, and only those satisfying both conditions are passed onto the NSIT-violating training data set. Figure~\ref{fig:gamma} illustrates the distribution of the interference witness measure $\gamma_{n,k}$ [Eq.~\eqref{eq:def.LGI.gamma}] values for 100,000 pseudo-randomly generated probability vectors in the NSIT-violating data sets of $N \in \lbrace 2,4,6 \rbrace$.
	
\subsection{Supervised Learning and Probability Vector Classification}
Supervised learning is a type of machine learning algorithm that infers a function from labeled training data. Training data sets are typically composed of pairs in which an input object is assigned a desired output value. For our purposes, the supervised learning task corresponds to a classification task, and the inferred function is a classifier function $\mathcal{C}$, i.e., a map $\mathrm{out} = \mathcal{C}(\mathrm{in})$ between input objects (i.e., probability vectors) and output values (i.e., the state classification).
	
An example implementation of our proposed machine learning methodology is openly available in the Github repository listed as Ref.~\cite{GithubRep}, including a separate file documenting the statistical distributions underlying the pseudorandom generation of the coefficients $p'_m(n,k)$ for both training data sets. The code provided in this repository is written in Mathematica 13 \cite{Mathematica13}, and the supervised learning task is performed by Mathematica's built-in ``Classify[]'' function.\footnote{A comprehensive documentation of the machine learning techniques available in Mathematica is provided in Ref.~\cite{B:21} and accessible online at \href{https://www.wolfram.com/language/introduction-machine-learning/}{https://www.wolfram.com/language/introduction-machine-learning/}.}\ In our example implementation, the training data set is generated such that all probability vectors from the NSIT-conforming [NSIT-violating] data set are assigned the desired output value 0 [1]. The resulting classifier function then takes an $N(N^2-1)$-dimensional probability vector as input and returns either 0 or 1 based on whether it has determined the state associated with the input vector to be of the NSIT-conforming or the NSIT-violating type. As a sanity check and to test the robustness of the classifier, we can feed the classifier function probability vectors for which the classification is known \textit{a priori} [e.g., through independent manual determination of the NSIT condition based on the spectral decomposition of Eq.~\eqref{eq:FP.SpecDecomp} or via the Bloch sphere inequality for emergent probabilities, e.g., Eq.~\eqref{eq:BlochSphere.Ineq.EmProb} for the single-qubit case] and evaluate its performance based on the accuracy of its output classifications. 
	
We stress that the sole intention of the provided code is to serve as a proof-of-principle implementation for our proposed machine learning methodology. As such, several refinements and extensions will be required in order to model real experimental applications and/or realistic large-$N$ systems.
	
\section{Conclusions} \label{sec:C}
Starting from the density operator form of the Schrödinger equation [Eq.~\eqref{eq:SchrEq}], we derive a formally equivalent probability vector representation [Eq.~\eqref{eq:SchrEq.FPform}] which describes the quantum dynamics of a system in terms of the probabilities associated with its observables. Our analysis demonstrates that the probability vector representation is uniquely suited to study features in the evolution of general probabilistic systems that are relevant with respect to the NSIT condition [Eq.~\eqref{eq:def:NSIT}] and its possible violation. Due to the specific form of Eq.~\eqref{eq:SchrEq.FPform.SolExpForm}, an exact diagonalization of the Hamiltonian is not required in this formalism, which has many advantages when $N$ is large.
	
After a measurement has been performed, the postmeasurement density operator may no longer be positive semidefinite (as evidenced by the fact that its spectrum may contain negative eigenvalues), which is indicative of NSIT violations and a dynamical evolution that cannot be understood classically (in the sense of being compatible with MR). While the negativity of quasiprobabilities has previously been considered as an indicator of quantumness in the Wigner-Weyl representation \cite{Ho:17,HBLO:19}, the difference in our approach is that all initial probabilities are always non-negative for both classical and quantum dynamics. This allows us to pinpoint what physical consequences the requirement of no backreaction/interference that is encoded in the NSIT condition entails. We also note positive operator-valued measure (POVM) based \cite{MKCM:15} and resource-theoretic \cite{MC:19} approaches to modeling LGIs and quantifying invasiveness, respectively.
	
The extent to which the NSIT condition is violated by quantum dynamics is ultimately determined by the evolution of probability distributions associated with the observables of the system under consideration and can be quantified using the interference witness measure defined in Eq.~\eqref{eq:def.LGI.gamma}. For single-qubit systems, we derive its explicit relationship to the backreaction of a measurement [Eqs.~\eqref{eq:SQ.gamma.a}, \eqref{eq:SQ.Delta.def}, and \eqref{eq:DeltaMeasure}].
	
As motivated by our argumentation in Sec.~\ref{sec:PVR.SE}, we expect our scheme to be more efficient computationally compared to conventional equation-solving approaches, particularly for the dynamics of large-$N$ systems. In this regime, the exploration of machine learning techniques (especially big data methods) appears to hold a lot of promise. The explicit treatment of large-$N$ systems (e.g., $N \simeq 2^{10^3}\!\mathrm{-}2^{10^4}$ in condensed matter systems) as well as possible experimental realizations of NSIT tests (for instance in quantum Hall systems) will be considered in future works. \\
	
\section*{Acknowledgements}
We would like to thank the anonymous referees for helpful comments and useful suggestions. This work was supported by an Okinawa Institute of Science and Technology (OIST) SHINKA Grant. MH is supported by Grant-in-Aid for Scientific Research (Grant No.\ 21H05188, No.\ 21H05182, and No.\ JP19K03838) from the Ministry of Education, Culture, Sports, Science, and Technology (MEXT), Japan. SM is supported by the Quantum Gravity Unit of OIST and would like to thank the Particle Theory and Cosmology Group at Tohoku University for their hospitality over the course of his research visit.

\end{document}